\documentclass[preprintnumbers,superscriptaddress,showpacs,pra]{revtex4}%
\usepackage[colorlinks=true,linkcolor=blue]{hyperref}
\usepackage{amssymb}
\usepackage{amsfonts}
\usepackage{amsmath}
\usepackage{graphicx}%
\providecommand{\U}[1]{\protect\rule{.1in}{.1in}}

\let\stdsection\section
\renewcommand\section{\nopagebreak\stdsection}
\begin{document}
\title{On "Quantum interference with slits" and its "revisited"}
\author{J. C. Ye, Y. Li, Q. Chen, S. G. Chen, and Q. H. Liu}
\affiliation{School for Theoretical Physics, School of Physics and Electronics, Hunan
University, Changsha 410082, China}
\date{\today}

\begin{abstract}
Marcella in 2002 published a "quantum-mechanical" treatment of the famous
single- and double-slit interference experiment in classical wave optics by a
simple assumption that quantum mechanical wave function is a constant anywhere
within a slit. Rothman and Boughn in 2011 commented that Marcella introduced
no quantum physics into the problem other than\ a symbol substitution $p=\hbar
k$, and the used essentially the classical wave optics. In present comment, we
point out that though Marcella made a fundamental mistake, the problem is
nevertheless remediable to give the satisfactory quantum mechanical results
which are in qualitatively agreement with experimental ones with material particles.

\end{abstract}
\maketitle
\preprint{REV\TeX4-1 }

In order to get a suggestive understanding of the famous uncertainty relation
$\Delta x_{i}\Delta p_{x_{i}}\simeq h$ ($i=x,y,z$), it is a typical treatment
in the quantum mechanics textbooks that use the well-known single- and
double-slit interference experiment in classical wave optics to illustrate the
relation. This is advantageous because there has not yet been a consistent,
full quantum--mechanical treatment of the slit experiment. In a 2002 issue of
present journal, Marcella \cite{Marcella} reported a "quantum-mechanical"
treatment of slit experiment, which is now widely cited as a convincing
quantum mechanical method reproducing exactly the same expressions originally
given by the classical wave optics, and the number recorded by the Google
scholar citation up to today (Jan. 24, 2019) is $24$ and $7$ of them are done
after 2011. Unfortunately, Marcella by and large misunderstood both his
derivations and results, as \textit{correctly} pointed out by Rothman and
Boughn in a 2011 paper published also in present journal, which attracts less
attention and the number recorded by the Google scholar citation is only two.
We agree with Rothman and Boughn on their comments on the Marcella's paper:
"while Marcella's procedure is useful in giving students practice with the
Dirac formalism, it has introduced no quantum physics into the problem other
than setting $p=\hbar k$, and has implicitly made all the assumptions that
show this is indeed a problem of classical optics." \cite{RothmanBoughn} The
situation is even worse for both the original paper \cite{Marcella} and its
comment \cite{RothmanBoughn} fail to point out that the wave function Marcella
introduced at very beginning of his treatment is incompatible with quantum
mechanics. Based on this observation, we are going to show that Marcella's
approach contains a truly quantum mechanical treatment with slits, though
qualitative but sufficient. We focus on the single-slit experiment in present
comment for sake of the simplicity, and the same manner can be directly used
to deal with the multi-slit one.

It is accepted that from the well-known single- and double-slit interference
experiment in classical wave optics, we can really reach the result $\Delta
x_{i}\Delta p_{x_{i}}\simeq h$ with an Einstein relation $p=h/\lambda$. To see
it, let us consider a setup in Fig.1.

The diffraction angle $\theta$ is drawn from the centre of the
single slit of openness $a$. In classical wave optics, a plane wave of wave
length $\lambda$ is initially incident to the plane $z=0$, and can be slightly
diffracted to a small angle $\theta$ after through the slit, the diffraction
amplitude on the distant screen is proportional to
\ \cite{Marcella,RothmanBoughn}
\begin{equation}
\frac{\sin\left(  a\pi\sin\theta/\lambda\right)  }{a\pi\sin\theta/\lambda
}.\label{1}%
\end{equation}
The first minimum of $\sin\left(  a\pi\sin\theta_{\min}\text{/}\lambda\right)
$ occurs at
\begin{equation}
\frac{a\pi\sin\theta_{\min}}{\lambda}=\pm\pi,\text{i.e., }\sin\theta_{\min
}=\pm\frac{\lambda}{a}.\label{2}%
\end{equation}
Now, let the incident plane wave represent a beam of matter particles of
momentum $p$, the momentum uncertainty along the direction of measuring the
width $a$ of slit is%
\begin{equation}
\delta p_{y}\simeq\left\vert p\sin\theta_{\min}\right\vert =p\frac{\lambda}%
{a}.\label{3}%
\end{equation}
Since the position uncertainty through the slit can be $\delta y\simeq a$, we
have then with $p=h/\lambda$,
\begin{equation}
\delta y\delta p_{y}\simeq\lambda p=h.\label{4}%
\end{equation}
Thus, in symbols of quantum mechanics, Eq. (\ref{1}) is%
\begin{equation}
\frac{\sin\left(  ap_{y}/2\hbar\right)  }{ap_{y}/2\hbar}.\label{5}%
\end{equation}
where $p_{y}\simeq p\sin\theta$, and $\theta\sim0$ for guaranteeing the
validness of the paraxial approximation.

The key step of Marcella's work was a seemingly reasonable assumption that the
wave function of the particle is along $y$-axes \cite{Marcella}
\begin{equation}
\psi(y)=1/\sqrt{a},\text{ for }y\in\text{slit, and }\psi(y)=0,\text{ for
}y\notin\text{slit. }\label{6}%
\end{equation}
The corresponding wave function in momentum space is%
\begin{equation}
\phi(p_{y})=\sqrt{\frac{a}{2\pi\hbar}}\frac{\sin\left(  ap_{y}/2\hbar\right)
}{ap_{y}/2\hbar}\label{7}%
\end{equation}
which differs from result (\ref{5}) only by a proportional factor. The
diffraction pattern on the screen, given by $\left\vert \phi(p_{y})\right\vert
^{2}$, is the momentum space image of the slit represented by (\ref{6}) in
position space. However, according to quantum mechanics, the wave function
(\ref{6}) is fundamentally flawed for the kinetic energy is divergent for we
have%
\begin{equation}
\int_{-\infty}^{\infty}p_{y}^{2}\left\vert \phi(p_{y})\right\vert ^{2}%
dp_{y}=\frac{2\hbar}{\pi a}\int_{-\infty}^{\infty}\sin^{2}\left(  \frac
{ap_{y}}{2\hbar}\right)  dp_{y}\rightarrow\infty\text{.}\label{8}%
\end{equation}
If it is true, the particle would get an infinitely large energy via the slit
after through it, which is ridiculous. The classical wave optics suggests a
relation $\delta x\delta p_{x}\simeq h$, but the "quantum mechanics" leads to
a meaningless uncertainty relation $\Delta x\Delta p_{x}\rightarrow\infty$
instead of $\Delta x\Delta p_{x}\simeq h$ that is what Marcella implicitly
assumed to be true for sure but he actually failed to achieve. \cite{Marcella}
So, the wave function (\ref{6}) Marcella \cite{Marcella} introduced is not a
quantum mechanical object but a classical wave, as Rothman and Boughn
commented. \cite{RothmanBoughn}

In the following, we try to fix the problem associated with the wave function
(\ref{6}) and resort to a simple quantum mechanical model of the slit. Note
that the slit serves as nothing but a configurational confinement on the
motion along $y$-direction. According to quantum mechanics, there must be a
minimum energy coming from the motion in the $y$-direction along the slit. The
particle before the slit is $E_{0}=p^{2}/2\mu$, and after the slit, the energy
conservation requires%
\begin{equation}
E_{0}=E_{y}+\frac{p_{z}^{2}}{2\mu},\label{9}%
\end{equation}
where $E_{y}$ stands for permissible energies due to the slit confinement
placed along $y$-direction, which must have a minimum value, and the motion
along $z$-direction is free between the slit and screen and its energy
spectrum is continuous. A rough approximation, without loss of the physics
content, is to treat motion along the slit as an infinitely deep well, and the
ground state in it is simply%
\begin{equation}
\psi(y)=\sqrt{\frac{2}{a}}\sin\left(  \frac{\pi\left(  y+a/2\right)  }%
{a}\right)  .\label{10}%
\end{equation}
If the particle has mass $\mu$, the minimum energy is therefore%
\begin{equation}
E_{y}=\frac{1}{2\mu}\left(  \frac{\pi\hbar}{a}\right)  ^{2}.\label{11}%
\end{equation}
The wave function of (\ref{10})\ in momentum space is \ \cite{zhang,Tannoudji}%
\begin{equation}
\varphi(p_{y})=2\sqrt{\frac{a\pi}{h}}\frac{\cos\left(  ap_{y}/2\hbar\right)
}{\left(  \pi^{2}-a^{2}p_{y}^{2}/\hbar^{2}\right)  }.\label{12}%
\end{equation}
The first minimum of $\cos\left(  ap_{y}/2\hbar\right)  $ occurs at $p_{y\min
}(a/2\hbar)=\pm\pi/2$, i.e., $p_{y\min}=\pm\pi\hbar/a$, which differs from
that given (\ref{7}) by a factor $3/2$. The maxima beyond the zeroth one,
given by (\ref{12}), are much less appreciable than those given by (\ref{7}).
It is worth stressing that in the real single slit experiment with a beam of
material particles such as $C_{70}$ or $C_{60}$ molecules, \cite{C70,C60} the
diffraction patterns do not exhibit the presence of the maxima beyond the
zeroth one. The diffraction patterns on the screen are plotted in Fig.2.

Three results above (\ref{10})-(\ref{12}) are from purely quantum mechanical
considerations. They are very important in the following three aspects. 1, No
particle passes through the slit once $E_{y}\prec\left(  \pi\hbar/a\right)
^{2}/2\mu$ for there is no state of energy less than this one. Note that the
state of energy $E_{y}=0$ is prohibited by quantum mechanics. 2, The quantum
mechanics gives the exact result for the $y$-momentum uncertainty, $\Delta
p_{y}=\pi\hbar/a$, which can also be estimated from result (\ref{11}). 3, The
quantum mechanics gives the $y$-position uncertainty $\Delta y\simeq0.18a$, so
we have the quantum mechanical uncertainty relation $\Delta y\Delta
p_{y}\simeq0.56\hbar\succ\hbar/2$, completely compatible with $\delta y\delta
p_{y}\simeq h$ suggested by the classical wave optics. Therefore, although we
do not know dynamical details how the slit makes the propagation direction of
the particle slightly deviated from the original direction, the single slit
diffraction pattern can be understood within quantum mechanics. 

In conclusion, Marcella's approach contains a truly quantum mechanical
treatment with slits, with the erroneous wave function be replaced by the
ground state of the infinitely deep well modelling the slit.

\begin{acknowledgments}
This work is financially supported by National Natural Science Foundation of
China under Grant No. 11675051.
\end{acknowledgments}

\begin{figure}[h]
\includegraphics[height=9cm]{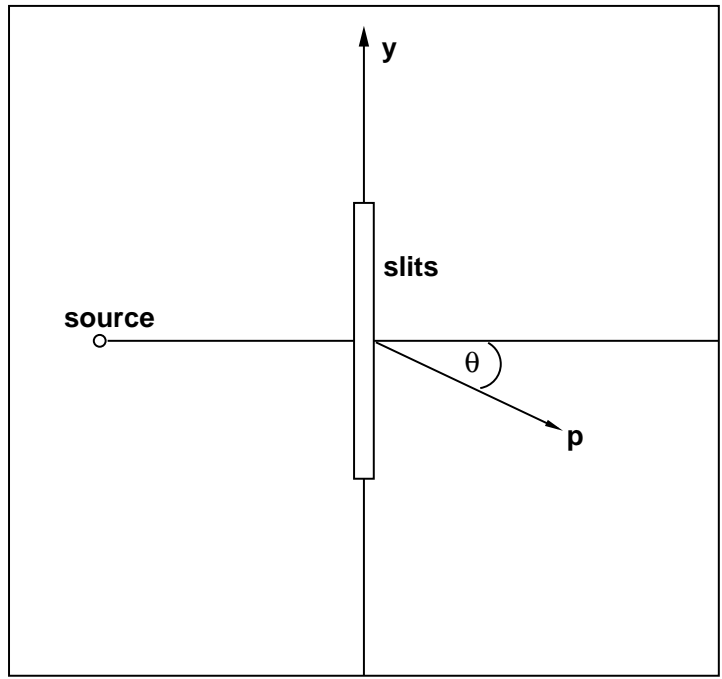}\caption{Marcella's setup for the one-
and two-slit experiment. After passing through the slit(s), the particle is
assumed to be scattered at an angle $\theta$ with a momentum $p$ and a
$y$-momentum of $p_{y}=p\sin\theta$. Note that i) the source is placed
sufficiently away from the slit such that the wave front near the slit is a
plane; and ii) the plane wave propagates along the $z$-axis and the slit is
situated on the plane $z=0$; and the diffracted angle $\theta$ is measured
from the centreline.}%
\label{fig}%
\end{figure}

\begin{figure}[h]
\includegraphics[height=9cm]{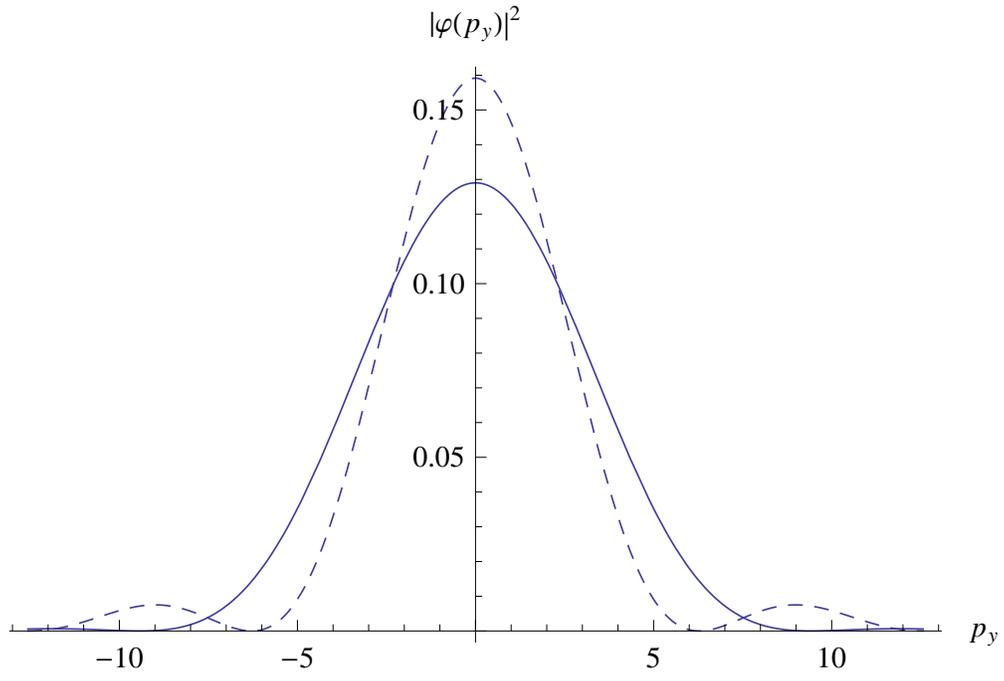}\caption{Diffraction distributions given
by (\ref{7}) (dashed line) and (\ref{12}) (solid line) respectively. In
quantum mechanics, the momentum uncertainty for dashed line is divergent but
in classical wave optics $\Delta p_{y}\simeq h/a$. }%
\end{figure}

\end{document}